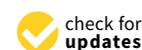

*Article*

# A Changing-Look AGN to Be Probed by X-ray Polarimetry

**Beatriz Agís-González** [1,2,*] ⬦, **Damien Hutsemékers** [1] and **Giovanni Miniutti** [3]

1 Space Sciences, Technologies and Astrophysics Research (STAR) Institute, Université de Liège, Allée du 6 Août 19c, B5c, 4000 Liège, Belgium; hutsemekers@astro.ulg.ac.be
2 Instituut voor Sterrenkunde, KU Leuven, Celestijnenlaan 200D, bus 2401 3001 Leuven, Belgium
3 Departamento de Astrofísica, Centro de Astrobiología (CSIC-INTA), Campus ESA-ESAC, Villanueva de la Cañada, 28691 Madrid, Spain; gminiutti@cab.inta-csic.es
* Correspondence: bagis@uliege.be

**Abstract:** Active galactic nuclei (AGN) produce the highest intrinsic luminosities in the Universe from within a compact region. The central engine is thought to be powered by accretion onto a supermassive black hole. A fraction of this huge release of energy influences the evolution of the host galaxy, and in particular, star formation. Thus, AGN are key astronomical sources not only because they play an important role in the evolution of the Universe, but also because they constitute a laboratory for extreme physics. However, these objects are under the resolution limit of current telescopes. Polarimetry is a unique technique capable of providing us with information on physical AGN structures. The incoming new era of X-ray polarimetry will give us the opportunity to explore the geometry and physical processes taking place in the innermost regions of the accretion disc. Here we exploit this future powerful tool in the particular case of changing-look AGN, which are key for understanding the complexity of AGN physics.

**Keywords:** AGN polarization; X-ray polarization; changing-look AGN

## 1. Introduction

Active galactic nuclei (AGN) are intrinsically the brightest non-transient objects in the Universe, producing very high luminosities from a concentrated volume. It is generally accepted that accretion onto a supermassive black hole (SMBH) is at the origin of this huge energy release that is widely spread across the whole electromagnetic spectrum, from $\gamma$-rays to radio. Thus, the term AGN includes a large variety of subtypes that collectively occupy a vast parameter space. One of the key ideas at the base of the current understanding of this disparate zoo is the unification model (UM) of AGN [1], for which the basic premise is the ubiquitous presence of an obscuring torus around the central engine, so that the observed diversity simply reflects different viewing angles of an axisymmetric geometry. Consequently, type-1 AGNs are observed with a direct view of fast-moving material close to the SMBH, resulting in broad emission lines in their optical/UV spectra, while type-2 AGNs are observed from a more edge-on view, intercepting the obscuring torus that blocks the emission of the broad line region (BLR) component from our line of sight (LOS). Actually, this transition between type 1 and type 2 is progressive with inclination, and as such we can find subtypes 1.2, 1.5, 1.8, and 1.9 [2]. The higher the subtype, the weaker the broad lines, up to subtype 1.9 where the broad component of H$\beta$ disappears while a weak broad H$\alpha$ component still remains.

This scheme was consolidated with the discovery three decades ago of polarized broad emission lines (PBLs) in the type-2 AGN—more specifically Seyfert 2 galaxy (Sy2)—NGC 1068 [3]. Since then, polarimetry has become an important tool in the study of AGN. This technique provides information on the physical structures that are below the resolution limit of the telescope [4]. Polarization can be only produced by some specific mechanisms, e.g., scattering by particles, which breaks the symmetry in the radiative source, allowing us to reconstruct the geometry of the unresolved scattering medium.

Optical and radio polarimetry constitute very powerful and sensitive tools for a wealth of astronomical sources, ranging from neutron stars to AGN. Nevertheless, the lack of polarimeters on-board space missions has prevented X-ray polarimetry from evolving in parallel with X-ray spectroscopy, photometry, or timing. This will change over the next decade with the scheduled launches of X-ray observatories equipped with the latest instrumentation sensitive to polarized light, e.g., IXPE [5,6] or eXTP [7]. The main goal of this manuscript is to show the potential of X-ray polarimetry, along with optical polarimetry, for the study of an intriguing and rare type of AGN: the changing-look (CL) AGN. The spectra of this class of objects changes from type 1 to type 2, or between the different subtypes, and vice versa, once or several times. Thus, they represent a violation of the UM, proving that type-1/2 classification is not only angle-dependent, and providing us with new insights into the physics and phenomenology of AGN.

Following that aim, in Section 2 we will introduce CL AGN and explain the functionality of optical polarimetry in that context. In Section 3, we will summarize our results for the changing-look Seyfert galaxy ESO 362-G18 and finally, in Section 4, we will show how X-ray polarimetry could give us decisive complementary information.

## 2. Changing-Look Scenario and Optical Polarimetry as a Diagnostic Tool

There exist a handful of well documented cases of rare events where the source transits from type-1 to type-2 states, or vice versa, i.e., its broad lines disappear or appear in its optical/UV spectrum. These are namely the Seyfert galaxies, NGC 3516 [8] NGC 7306 [9], NGC 4151 [10], Fairall 9 [11], Mrk 1018 [12,13], Mrk 993 [14], NGC 1097 [15], NGC 7582 [16], NGC 2617 [17] and Mrk 590 [18], and the quasars SDSS J015957.64+003310.5 [19], SDSS J101152.98+544206.4 [20,21], and SDSS J155440.25+362952.0 [22]. Interest in CL AGN has been growing in recent years, as evidenced by systematic searches of catalogs, (see e.g., [23–25]). Astronomers have proposed two main plausible physical processes to explain the origin of these transitions:

1. The AGN activity depends on the availability of gas to fuel the black hole. The mechanisms by which cold gas is transported from the galaxy disc at kpc scales to further down into the inner few parsecs to trigger AGN activity determine the efficiency with which the black hole is fueled [26]. Large changes in the accretion onto the SMBH can be responsible for these changing looks, either creating or disrupting the broad line region. This is because broad lines are formed by photoionization driven by emission from the accretion disc; then an intrinsic dimming of the continuum source reduces the number of photons available to ionize the gas, resulting in a net decrease of the emission line intensity. Actually, below a given critical accretion rate (or correspondingly luminosity), no BLRs can be formed nor sustained over a long time, thus those sources will always appear as type-2 objects [21,27,28].

2. On the other hand, if the inclination of the source is such that our LOS intercepts the upper or lower edge of the torus, rapid changes in the observed flux can occur when the obscuring dusty torus is considered as a patchy structure where moving clumps open or block a clear view of the central region of the AGN [29,30]. In this configuration, the disappearance and appearance of broad emission lines can be explained by variable obscuration.

AGN optical polarimetry comes into play as a diagnostic tool to distinguish between both phenomena: a variable accretion rate or variable obscuration. Specifically, a clear dichotomy between type-1 and type-2 AGN arises in broad-band imaging polarimetry (see e.g., [31] for quasars and [32] for Seyfert galaxies). Sy2s show high broad-band degrees of polarization, $p > 7\%$, while Seyfert 1 galaxies (Sy1s) exhibit low degrees, $p \leq 1\%$. Following the models initially developed by the authors of [32–34], the polarization of Seyfert galaxies can be interpreted by the scattering off two regions: an equatorial ring located inside the dusty torus at the origin of "parallel" polarization, and a more extended polar-scattering region placed along the symmetry axis at the origin of "perpendicular" polarization. Considering the inclination as the angle between the system axis and the LOS, Sy1s are seen at low inclinations, and then both scattering regions are intercepted in our LOS, resulting in a low polarization and a polarization position angle (PPA) parallel to the system axis. Sy2s present high inclinations, so that the equatorial scattering region is hidden by the obscuring torus and consequently highly polarized polar-scattered light is detected while the PPA becomes perpendicular to the system axis. Then, if the disappearance (appearance) of broad emission lines in CLs is caused by torus dust clouds hiding (not hiding) the Seyfert core—i.e., the continuum source, the BLR, and the equatorial scattering region—the high (low) polarization typical of Sy2s (Sy1s) will be expected. However, there exist a number of Sy1 galaxies where polar scattering dominates, as is the case for Sy2 galaxies [34]. The trends on p and PPA identified with PBL Sy2 are also plausible in polar-scattered Sy1 galaxies if the direct view of the AGN continuum passes through the upper layers of the dusty circumnuclear torus.

Relying on this premise, we used polarimetry to disentangle the origin of the change of look exhibited by the quasar J101152.98+544206.4 (hereafter J1011+5442) [20], for which broad emission lines disappeared between 2003 and 2015 [21], i.e., a transition from type 1 to type 2. If the source presents high polarization remaining in a type-2 state, the cause of the change is expected to be a dusty clump of the torus which blocks the continuum source, the BLR, and the equatorial scattering region in our LOS. Thus, radiation escapes from the AGN by scattering inside the polar outflows so that a large polarization degree can emerge. On the contrary, if the exhibited polarization is low even in a type-2 state, the plausible scenario is a rapid decrease in the rate of accretion onto the SMBH that becomes insufficient to sustain the BLR (and torus) [21]. In this case, the equatorial scattering region redirects photons towards the observer, decreasing the net polarization.

We performed linear polarization data of this source in 2017 (see [20]) and found that J1011+5442 was still in a faint state, but nonetheless its polarization was very low, typical of unobscured Sy1s. This result supports the variable accretion rate scenario as suggested by [21]. Afterwards, [35] reinforced the use of polarimetry as a diagnostic tool by modeling the optical continuum polarization expected from CL AGN in a transition from type 1 to type 2. The author carried out simulations for the two described scenarios, variable accretion rate and variable obscuration, using the Monte Carlo radiative transfer code STOKES [36,37]. His results show the very distinctive polarimetric features for both cases and fully agree with the conclusions obtained by [20] for J1011+5442.

## 3. The Changing-Look Seyfert Galaxy ESO 362-G18

Here we focus on a specific CL Seyfert galaxy: ESO 362-G18. This target is usually classified as Seyfert 1.5 (see e.g., [38] (EMMI data) or [39–41]). However, the authors of [42] retrieved a spectrum belonging to the Six-degree Field (6dF) Galaxy Survey which does not show the broad lines mandatory to classify the source as type 1. Indeed, [42] classified it as a naked, or true, Sy2 galaxy, i.e., with a dimmed continuum unable to sustain the BLR [43]. We also retrieved more spectropolarimetric data of ESO 362-G18 from the ESO archive taken with EFOSC1 in 2006 and were awarded new observations in polarized light from FORS2 in 2016. The four spectra (6dF, EMMI, EFOSC1 and FORS2) are shown in Figure 1. We carried out a rigorous analysis of this dataset applying single Gaussian fits for the narrow lines, double Gaussian fits for the broad H$\beta$ component, and triple Gaussian fits for the broad H$\alpha$ component necessary to give an account for the broad H$\alpha$ wings, (e.g., [38,44,45]). From our spectral analysis, we found ESO 362-G18 in a Sy1.9 configuration in 6dF and EFOSC1 data, while striking broad H$\alpha$ and H$\beta$ emission lines were detected in both EMMI and FORS2 data, leading to a Sy1.5 classification.

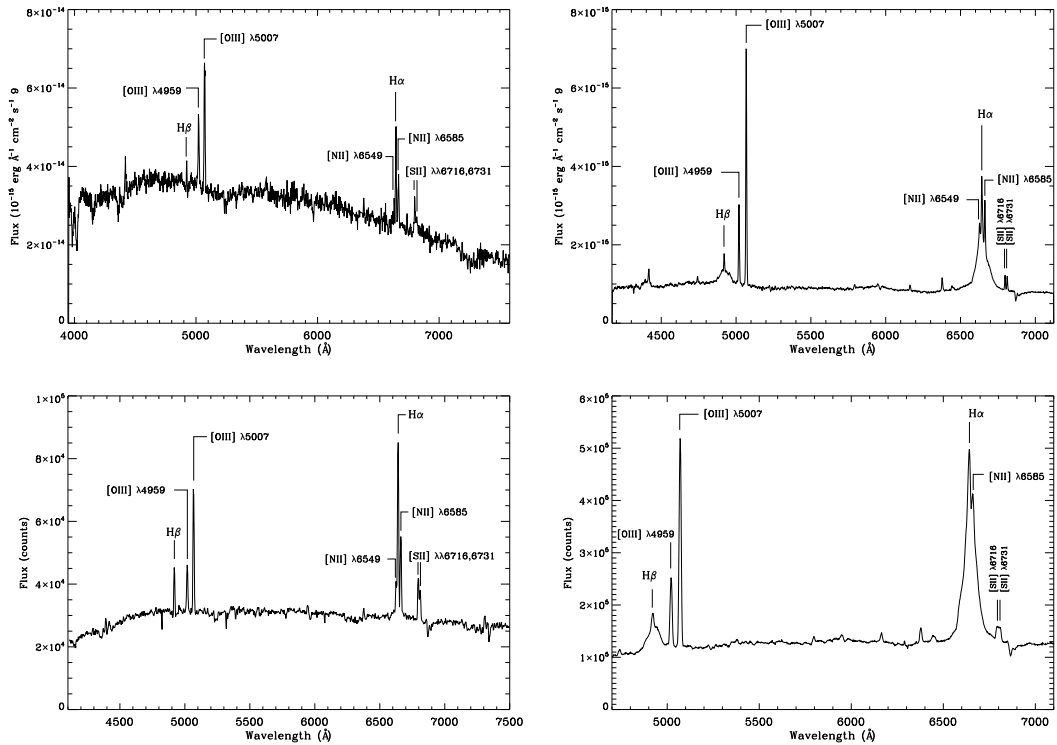

**Figure 1.** Four datasets of ESO 362-G18. From right to left and from top to bottom: 6dF spectrum dated 30 January 2003 [42] (Sy1.9); EMMI data dated 18 September 2004 [38] (Sy1.5); EFOSC1 data dated 21 September 2006 (Sy1.9); FORS2 data dated 29 March 2016 (Sy1.5).

According Figure 1, the broad lines of ESO 362-G18 appear in the 20 months between the first two datasets in a transition 1.9–1.5. Subsequently the broad lines disappear again, returning to type 1.9 two years later. Finally, after 10 years, the 1.5 configuration comes back again. Thus, we confirm ESO 362-G18 as a changing-look Seyfert galaxy with three changes of state. Both EFOSC1 and FORS2 data were secured in polarized light, allowing an analysis of a CL Seyfert galaxy in spectropolarimetry. These results, still in progress, will be published in Agís-González et. al (in prep.).

On the other hand, on the basis of the analysis carried out by [41], ESO 362-G18 is known to exhibit potential X-ray variability. Thus, we decided to start a parallel multi-epoch X-ray analysis to assess the X-ray flux and spectral variability of the source on timescales of days to years [46], between November 2005 and January 2011. In total, 45 X-ray observations were performed with Swift, XMM-Newton, Suzaku, and Chandra spacecraft, including two rounds of monitoring—one of them with Chandra with five observations over 2 weeks in May 2010, and another one with Swift with 36 observations over 2 months in November 2010–January 2011, for which results will be published in Miniutti et al. (in prep.).

The high-quality observations (53 ks with XMM-Newton in January 2010 and 41 ks with Suzaku in April 2008) enabled us to decompose the complex X-ray spectrum of ESO 362-G18 into its different components, finding among them an X-ray relativistic reflection component originating in the innermost regions of the accretion disc. Then, its spectral shape is distorted by the gravitational effects of the SMBH, which allow us to get valuable information about the central engine. Particularly, we measured an inclination between the disc axis and our LOS of $53° \pm 5°$. Such a high inclination is further supported by different estimates based on virial assumptions as well as on the relationship between black hole mass and stellar velocity dispersion (e.g., [47]) taking into account the stellar velocity dispersion measured by [48] for ESO 362-G18.

At the same time, we detected two X-ray absorption events in our dataset: one during a single XMM-Newton observation taken in January 2006 and another one during the Swift monitoring from November 2010 to January 2011. Swift and XMM-Newton carry on-board optical monitors, so we used these optical photometric data to check if both absorption events were also detected at UV wavelengths, which would imply dust content in the absorber. In that case, since the BLR is defined to be dust-free, i.e., within the dust-sublimation radius, the absorbing material would have to belong to the dusty torus, e.g., [49]. From our two XMM-Newton observations (January 2006 for which an absorption event was detected and January 2010, not absorbed), along with the single pointed Swift one (November 2005, not absorbed), we concluded that the UV emission is affected by the occultation event, which places the absorbing material in the dusty torus. On the other hand, during the Swift monitoring the UV emission keeps constant, however the detailed covering fraction evolution coupled with the derived ionization and column density enabled us to estimate the properties of this detected absorber. Actually, the cloud location was found to lie just at the dust sublimation radius, which may be the reason why no clear absorption-induced variability was seen in the UV range (Miniutti et al. (in prep.)). Furthermore, the derived high inclination is consistent with the idea that our LOS is grazing the obscuring torus (which has a typical half-opening angle of the order of 45° or so). If the torus is not homogeneous but clumpy, such a high inclination may intercept from time to time some of the clumps of the upper layers of the obscuring torus, possibly explaining why ESO 362-G18 exhibits the described changes of look. In addition, we did not detect any evidence for dramatic changes in the nuclear activity (this is demonstrated by the stability of the X-ray intrinsic luminosity over a few years), so the most natural explanation for the change of look is that intervening dusty absorbing structures are present in our LOS towards the BLR in the epochs when the source exhibits a Seyfert 1.9 optical spectrum, supporting the variable absorption scenario. Nevertheless, we remark that it is unlikely that we are observing exactly the same intervening structures in the optical and X-rays (Agís-González et al. (in prep.)).

Taking into account this set of results, we would have expected to find a higher polarization degree in the type-1.9 configuration of ESO 362-G18 (EFOSC1 data) and a lower polarization degree in the 1.5 state with a perpendicular PPA in both configurations, i.e., the typical spectropolarimetric features of polar-scattered Sy1s. However ESO 362-G18 is basically unpolarized. Although these are still preliminary results, the degree of polarization is very low and does not show a remarkable change between type-1.9 and type-1.5 data (see Table 1), while the value of the PPA is consistently constant and perpendicular in both observations, and broad emission lines do not appear in the corresponding polarized spectra. There is still work in progress, but our main conclusion is that the optical scattering regions are not efficient, and may be too small or not well placed. X-ray polarization will give us more information about this intriguing case which seems to also contradict the described standard polarization scheme.

**Table 1.** Preliminary values of the polarization parameters on ESO 362-G18 for both exhibited looks in the optical range.

|  | Type-1.9 | Type-1.5 |
| --- | --- | --- |
| p | $0.46 \pm 0.07\%$ | $0.22 \pm 0.02\%$ |
| ΔPPA | ~60° | ~70° |

## 4. X-ray Polarimetry as a Diagnostic Tool in the ESO 362-G18 Case

Simulations carried out by [50] show that the observed polarization dichotomy in the optical/UV band should extend into the X-ray range. These authors adopt a lamp-post geometry between 1 and 100 KeV and include X-ray reprocessing of the primary radiation by the accretion disc. Their results show that for all viewing directions below the torus horizon, i.e., type-2 AGN, high and perpendicular polarization arises, while towards a face-on view, i.e., type-1 AGN, electron scattering in the equatorial region dominates and produces a net parallel polarization at lower polarization degrees. Consequently, X-ray polarization can be used as a tool in the changing-look scenario as in optical polarimetry probing smaller scales.

In the particular case of ESO 362-G18, X-ray polarimetry could give us further information about the X-ray scattering region, which is supposed to be located the inner parts of the accretion disc, where the primary X-ray emission is reflected and reprocessed. We have already detected X-ray disc reflection in the analyzed X-ray observations. Hence, the X-ray scattering region should be detected and would add constraints to the existence or not of scattering regions in ESO 362-G18. It is worthwhile to highlight that in the X-ray domain, unlike the optical band, there will not be a dilution of the intrinsic polarization by the host galaxy, making the detection of the scattering regions easier.

Then, we could obtain two hypothetical results from X-ray polarimetry:

1.  No measured polarization in the X-ray domain. This fact means that the X-ray scattering region cannot be detected. We will have to estimate and establish a detection limit for the measured polarization to probe the absence of X-ray scattering regions.
2.  Otherwise, we could measure X-ray polarization as high or low depending on the state of the source during the observations, and get constraints about the X-ray scattering region. Probably this will help us to infer some explanation about the non-detection of the optical scattering regions and develop polarization models of CL AGN.

In order to give a rough assessment of the feasibility of observing ESO 362-G18 with a forthcoming X-ray polarimeter, we followed the work carried out by Marin et al. [51] who investigated the detectability of a typical type-2 AGN with IXPE by performing simulations. Since type-2 AGN are obscured, they are more difficult to detect than type 1 (unobscured). Due to their CL nature, we decided to consider the X-ray obscured states of ESO 362-G18 for this discussion. Those authors fixed a X-ray flux in the 2–8 KeV band of $5.19 \times 10^{-12}$ erg cm$^{-2}$s$^{-1}$ and a galactic column density towards the AGN of $2.99 \times 10^{20}$ cm$^{-2}$ for their simulations. Taking into account the best data for the X-ray absorbed state of ESO 362-G18, i.e., those performed with XMM-Newton on 28 January 2006 [46], the X-ray flux ($3.7 \times 10^{-12}$ erg cm$^{-2}$s$^{-1}$ in the 2–10 KeV band) and the absorbing hydrogen column density LOS's ($1.75 \times 10^{20}$ cm$^{-2}$) can be compared to the values used in the models developed by [51]. Those models also require the expected polarization of the source, which in turn depends on many morphological parameters. The authors [51] obtain that, for a typical type-2 AGN, a 20 Ms observation with IXPE would be necessary to achieve detection of the polarized emission. Thus, by translating these results to ESO 362-G18, taking into account the changes of look of our source, the absorbed state observed with XMM-Newton, which is characterized by a column density of 3–4 $\times 10^{23}$ cm$^{-2}$ (still not optically thick [52]), and the derived high inclination of ~53° [46], we extend the range of the required exposure time to 2–20 Ms, since these characteristics would increase the expected X-ray polarization of the source, thus decreasing the needed observing time.

## 5. Conclusions

X-ray polarimetry gives us the opportunity to study AGN scattering regions in a unique way and complements the information on the innermost regions of the accretion disc and reprocessing sites provided by X-ray spectroscopy and timing, shedding light on the physical processes and geometry of these unresolved objects. Specifically, studying the inner parts of changing-look AGN will help us to probe in depth the AGN unification scheme, which seems to need additional components that only exceptional objects such as changing-look AGN can give us.

**Author Contributions:** B.A.-G. has led this work and prepared the manuscript. D.H. has contributed to the methodology and analysis of the optical polarization of ESO 362-G18, while G.M. has contributed to the methodology and analysis of the X-ray data. D.H. and G.M. have read the manuscript and contributed comments.

**Acknowledgments:** This work has been carried out thanks to the funding provided by the KU Leuven to B.A.-G. D.H. is a Senior Research Associate F.R.S.-F.N.R.S. The authors would like to acknowledge the two anonymous referees for their useful suggestions and Frédéric Marin for his comments on X-ray simulations which improved the quality of the paper.

**Conflicts of Interest:** The authors declare no conflicts of interest.

## Abbreviations

The following abbreviations are used in this manuscript:

| | |
|---|---|
| AGN | active galactic nuclei |
| BLR | broad line region |
| CL | changing-look |
| LOS | line of sight |
| PBL | polarized broad line |
| p | degree of polarization |
| PPA | polarization position angle |
| SMBH | supermassive black hole |
| Seyfert 1 galaxy(-ies) | Sy1(s) |
| Seyfert 2 galaxy(-ies) | Sy2(s) |
| UM | unified model |